\documentclass[fleqn,12pt,twoside]{article}
\usepackage{espcrc1}

\usepackage{epsf}

\def\be{\begin{equation}}
\def\ee{\end{equation}}
\def\bea{\begin{eqnarray}}
\def\eea{\end{eqnarray}}
\def\simgt{\,\rlap{\lower 3.5 pt\hbox{$\mathchar \sim$}}\raise 1pt \hbox {$>$}\,}
\def\simlt{\,\rlap{\lower 3.5 pt\hbox{$\mathchar \sim$}}\raise 1pt \hbox {$<$}\,}

\title{
\vspace*{-60pt}
{\normalsize \hfill {\sf UTHEP-462}} \\
{\normalsize \hfill {\sf UTCCP-P-129}} \\
{\normalsize \hfill {\sf September 2002}} \\
\vspace*{30pt}
Recent lattice results relevant for heavy ion collisions
\thanks{Talk presented at 
	the XVI International Conference on Ultrarelativistic Nucleus-Nucleus
	Collisions (Quark Matter 2002), July 18--24, 2002, Nantes, France}
}

\author{Kazuyuki Kanaya\address[UT]{Institute of Physics, 
        University of Tsukuba, \\ 
        Tsukuba 305-8571, Japan}%
}
       
\begin{document}

% typeset front matter
\maketitle

\begin{abstract}
I summarize recent lattice results on QCD at finite temperatures and densities.
Studies on the nature of the QCD transition at the physical point, continuum extrapolations of thermodynamic quantities, and new calculations of hadronic spectral functions applying the maximum entropy method are discussed. Recent advances in finite density QCD are also reviewed.
\end{abstract}

%%%%%%%%%%%%%%%%%%%%%%%%%%%%%%%%%%%%%%%%%%%%%%%%%%%%%%%%%%%%%%%%%%%%%
\section{INTRODUCTION}

Lattice QCD provides us with the most systematic way to calculate non-perturbative hadronic quantities directly from the first principles of QCD, without resorting to models and effective theories.
Because a simulation of the fully realistic case is still difficult, we have to carry out several extrapolations to extract physical quantities of phenomenological relevance. 
These include the continuum extrapolation, the extrapolation to the physical quark mass point, and, for heavy-ion collisions, the extrapolation to finite quark densities also. 
Recent progress in the computer power and simulation techniques have enabled us to put well-controlled systematic errors to several quantities. 

In this article, I review the status of lattice QCD at finite temperatures and densities, concentrating on studies which appeared after the Quark Matter 2001 conference \cite{reviews}. 
In Sec.~\ref{sec:T}, I summarize the developments in finite temperature QCD, selecting the topics of the approaches toward the physical point and the continuum limit. 
I also discuss recent studies of hadronic spectral functions adopting a new calculation technique, the maximal entropy method.
A noticeable trend of lattice studies in the last 1.5 years is a rapid increase of investigations at finite densities, induced by a study of QCD at small chemical potentials by Fodor and Katz \cite{FK}.
I introduce these developments in Sec.~\ref{sec:mu}. 
Short conclusions are given in Sec.~\ref{sec:summary}.

%%%%%%%%%%%%%%%%%%%%%%%%%%%%%%%%%%%%%%%%%%%%%%%%%%%%%%%%%%%%%%%%%%%%%
\section{FINITE TEMPERATURE QCD}
\label{sec:T}

%%%%%%%%%%%%%%%%%%%%%%%%%%%%%%%%%%%%%%%%%%%%%%%%%%%%%%%
\subsection{Toward the physical point}
\label{sec:3flavor}

\begin{figure}[t]
\begin{minipage}[t]{75mm}
\hspace*{1mm}
\centerline{(a) \epsfxsize=7.2cm \epsfbox{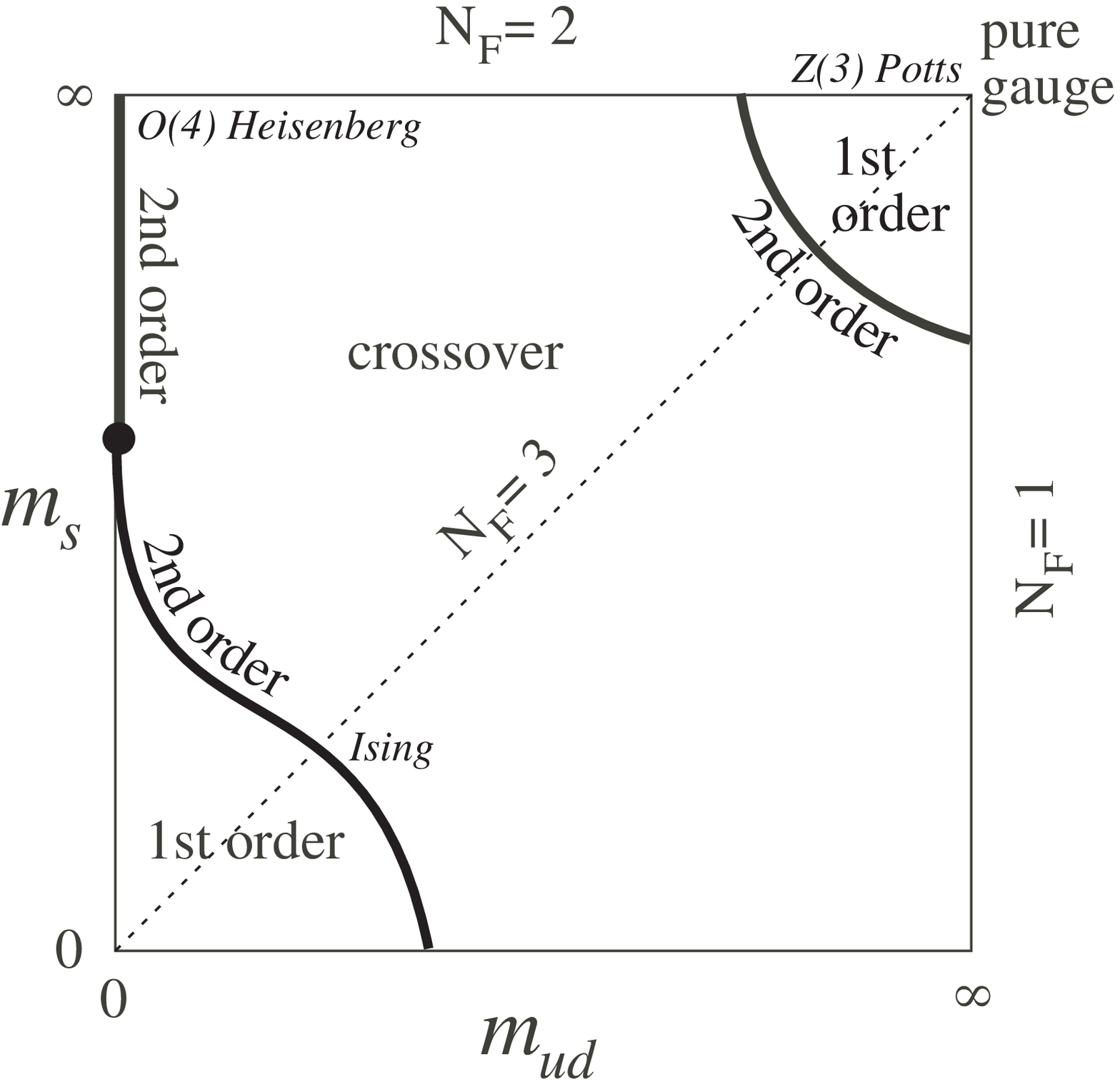}}
\end{minipage}
\hfill
\begin{minipage}[t]{77mm}
\centerline{(b) \epsfxsize=7.5cm \epsfbox{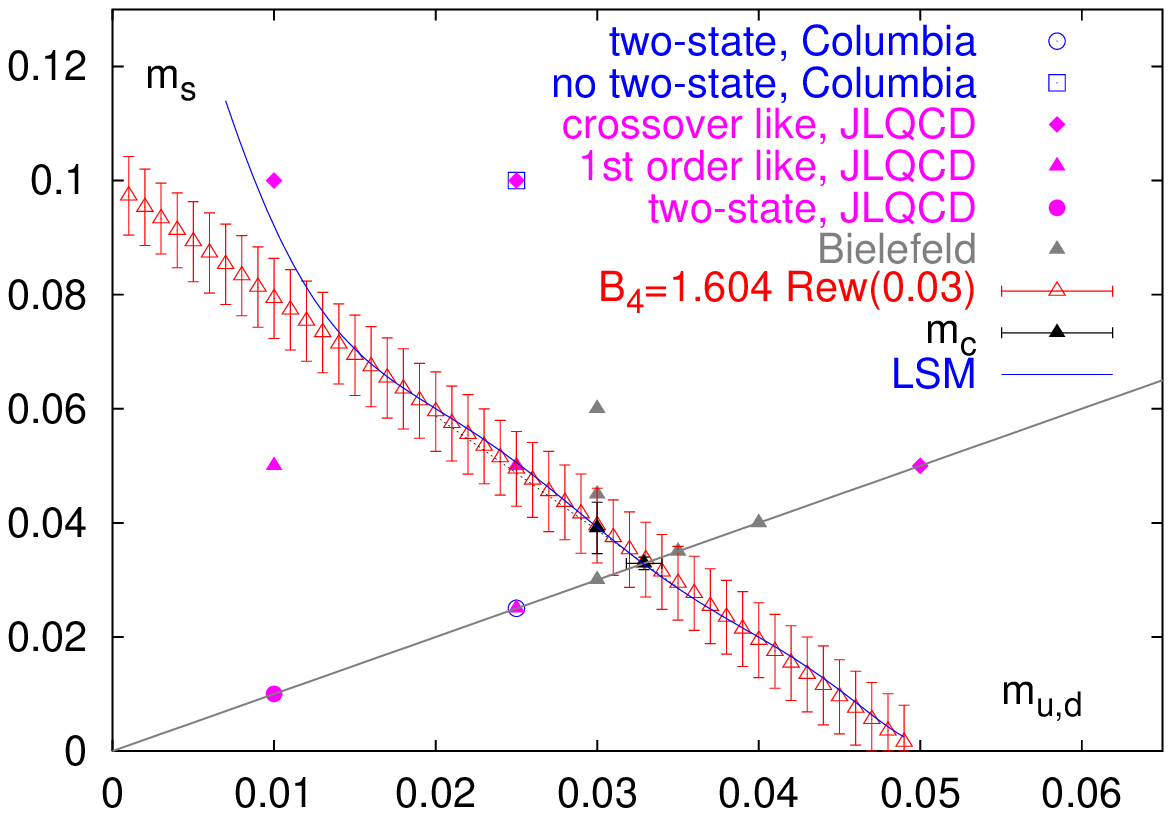}}
\end{minipage}
\vspace{-0.6cm}
\caption{Nature of the QCD transition as a function of quark masses.
(a) Theoretical expectations from various effective models.
(b) Phase structure around the three flavor critical point from a staggered quark simulation using the standard action on an $N_t=4$ lattice \protect\cite{BiSwLat02}.}
\label{fig:2+1}
\end{figure}

Because the cost to simulate dynamical quarks is quite high on the lattice, most large scale simulations are done in the approximation that only degenerate $u$ and $d$ quarks are treated dynamically ($N_F=2$ QCD).
However, since $m_s \sim 100$ MeV is around the transition temperature $T_c \sim 150$ MeV, inclusion of the dynamical $s$ quark is important ($N_F=2+1$ QCD).
Fig.~\ref{fig:2+1}(a) summarizes our expectations on the nature of the QCD transition as a function of quark masses based on various theoretical studies of effective models.

In the limit of heavy $s$ quark ($N_F=2$ QCD), the chiral transition at $m_{ud}=0$ is expected to be second order in the universality class of the three dimensional O(4) Heisenberg model \cite{PisarskiWilczek}.
Previous studies of two-flavor QCD have been done mostly on $N_t=4$ and 6 lattices.
Through improvement of the lattice theory, several quantities, such as the transition temperature and the EOS, have started to show convergence among different lattice formulations \cite{CPPACS01,Bi01}.
On the other hand, the expected O(4) scaling could be seen only with improved Wilson-type quarks \cite{Iwasaki97,CPPACS01}, and so far not with staggered quarks \cite{KL94,JLQCD98,MILC00}.

When we decrease $m_s$ along the $m_{ud}=0$ line, we will encounter a tri-critical point $m_s^*$ where the second order chiral transition turns into a first order transition, because the chiral transition of $N_F=3$ QCD at $m_{ud}=m_s=0$ is expected to be first order.
For $m_s < m_s^*$, the second order edge of the first order region will deviate from the vertical axis according to $m_{ud} \propto (m_s^{*} - m_s)^{5/2}$.
% What was not predicted definitely from effective theories is the location of the physical point. 
In phenomenological studies of heavy-ion collisions, it is important to know the location of the physical point in Fig.~\ref{fig:2+1}, for which a definite prediction is not available from effective theories. 

Identification of the physical point is not easy on the lattice either because it requires a precise scale determination at small quark masses in $N_F=2+1$ full QCD. 
Previous studies using staggered quarks suggest that the physical point locates in the crossover region \cite{Columbia90,JLQCD99}, while a study using Wilson quarks suggests a first order transition \cite{Tsukuba96}.
These simulations are done on $N_t=4$ lattices using the standard lattice actions.
Because lattice artifacts are expected to be large in these simulations, studies at larger $N_t$ using improved actions are called for to draw a definite conclusion.

At the Lattice 2002 conference, several new $2+1$ flavor simulations were reported.
The Bielefeld-Swansea Collaboration and the Columbia group presented updates of the staggered quark phase diagram at $N_t=4$ using the standard action \cite{BiSwLat02,ColumbiaLat02}.
The locus of the second order line is determined around the $N_F=3$ critical point using the Binder cumulant method (see Fig.~\ref{fig:2+1}(b)).
The results are consistent with the previous studies with standard staggered quarks at $N_t=4$. 

\begin{figure}[t]
\vspace{-2.5cm}
\hspace*{1mm}
\begin{minipage}[t]{76mm}
\centerline{\epsfxsize=7.5cm \epsfbox{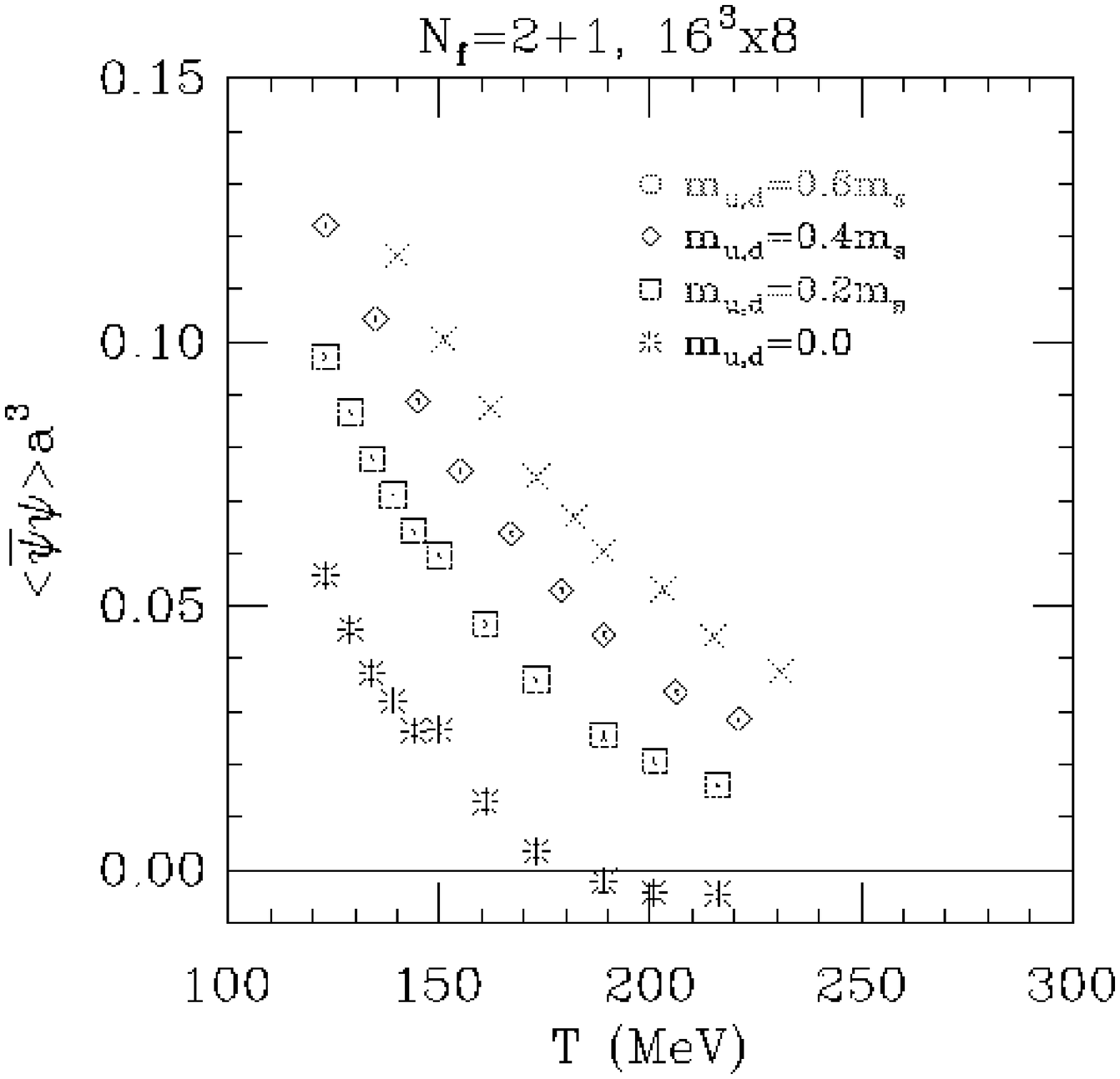}}
\end{minipage}
\hfill
\begin{minipage}[t]{76mm}
\centerline{\epsfxsize=7.5cm \epsfbox{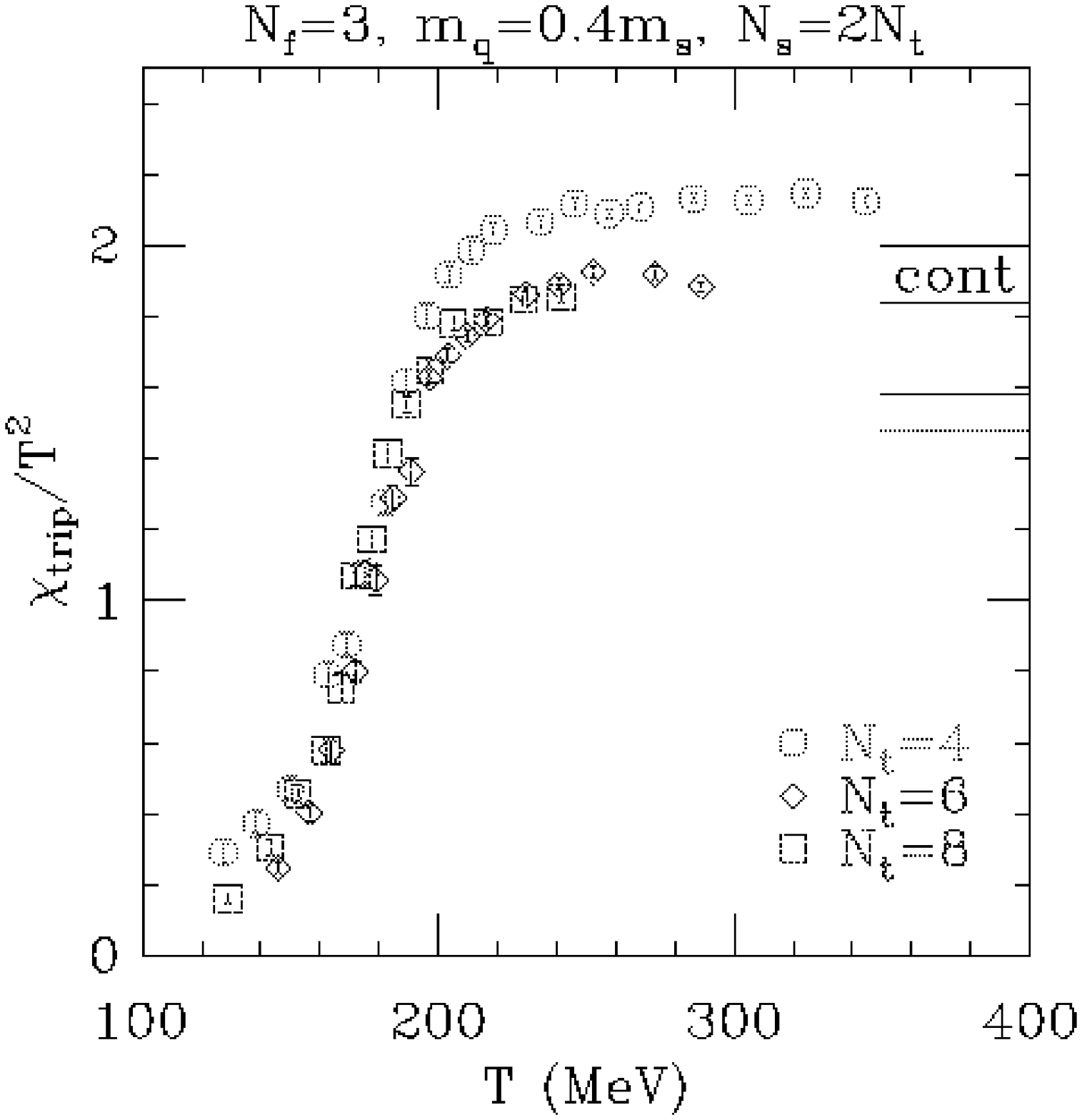}}
\end{minipage}
\vspace{-2.8cm}
\caption{$N_F=2+1$ and 3 staggered quark simulations using improved actions \protect\cite{MILCLat02}.
(a) Chiral order parameter at $N_t=8$ in $N_F=2+1$ QCD. The bursts are linear extrapolations in $m_{ud}$ to the chiral limit.
(b) Triplet quark number susceptibility as a function of $N_t$ in 3-flavor QCD.}
\label{fig:MILC}
\end{figure}

The MILC Collaboration presented results with staggered quark on $N_t=6$ and 8 lattices ($12^3\times6$ and $16^3\times8$) using improved actions (1-loop Symanzik gauge action and Asqtad staggered quark action) \cite{MILCLat02}.
Fixing $m_s$ to the physical $s$ quark mass value, three light quark masses were studied ($m_{ud}/m_s = 0.6$, 0.4, 0.2). 
They also performed $N_F=3$ simulations at $m_q/m_s^{phys} = 1$, 0.6 and 0.4.
Fig.~\ref{fig:MILC}(a) shows the quark condensate as a function of $T$ and $m_{ud}$. 
Because of large $m_{ud}$, the transition is crossover at all simulation points. 
A crude extrapolation to $m_{ud}=0$ shown in Fig.~\ref{fig:MILC}(a) suggests that the transition temperature is larger than about 170 MeV there, while, without data at small $m_{ud}$, it is difficult to conclude about the nature of the transition at the physical point. 

%%%%%%%%%%%%%%%%%%%%%%%%%%%%%%%%%%%%%%%%%%%%%%%%%%%%%%%
\subsection{Toward the continuum limit}
\label{sec:continuum}

Numerical data obtained on finite lattices must be extrapolated to the continuum limit to extract predictions for the real world. 
In finite temperature QCD, this corresponds to the limit of large $N_t$ because the temperature in physical units is given by $T=1/N_t a$.
On the other hand, the spatial lattice size $N_s a$ should be much larger than $T^{-1}=N_t a$ to approximately realize the thermodynamic limit. 
From experiences in quenched QCD, $N_s/N_t$ \simgt 4 is said to be safe for the EOS.
The computation is, therefore, quite expensive at large $N_t$.
This is a reason that major full QCD simulations have been limited to $N_t$ \simlt 6.

\begin{figure}[t]
\vspace{-0.5cm}
\begin{minipage}[t]{74mm}
\centerline{\epsfxsize=7.4cm \epsfbox{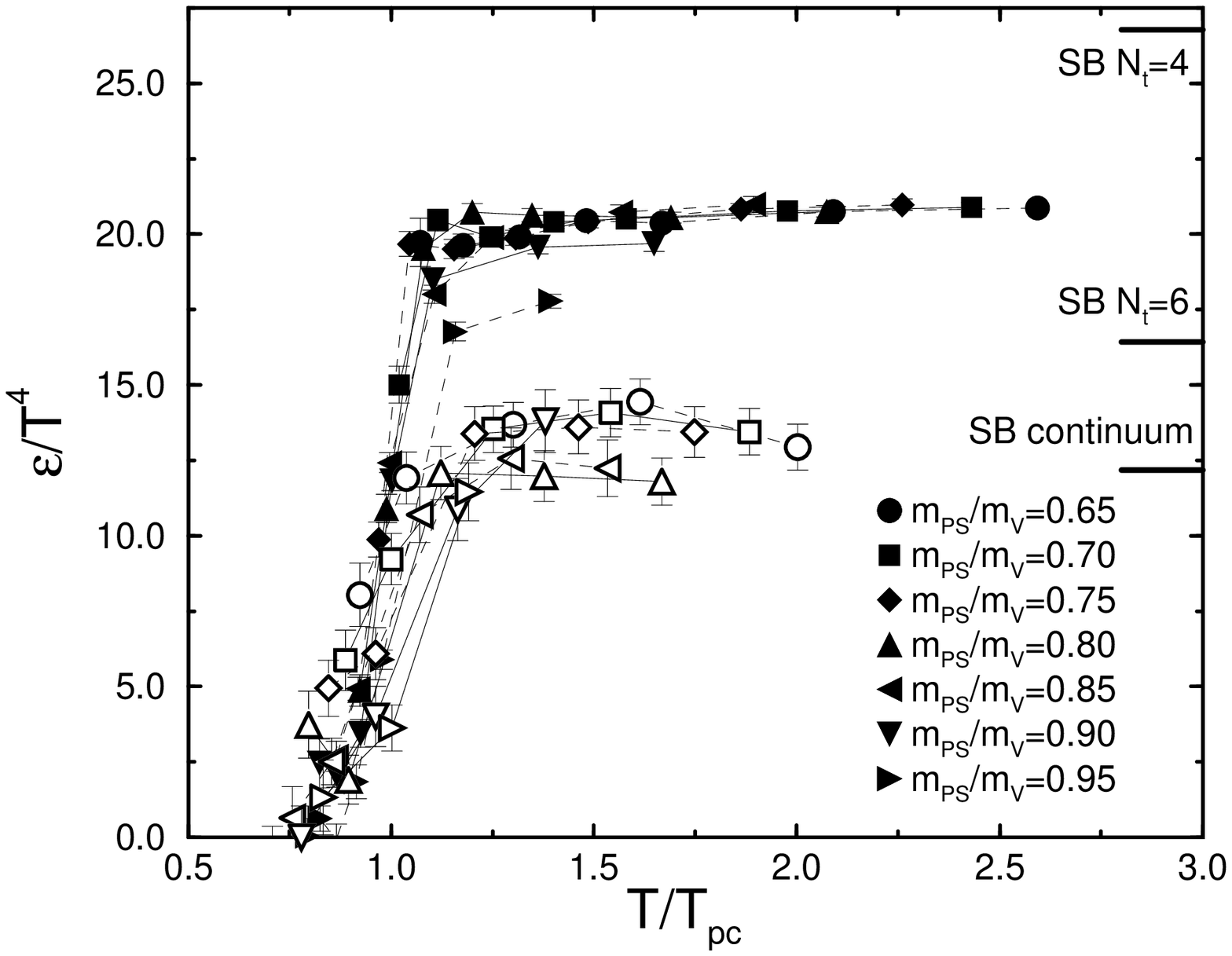}}
\vspace{-0.9cm}
\caption{The EOS from QCQ with Wilson-type quarks on $N_t=4$ (filled symbols) and 6 (open symbols) lattices using improved actions \protect\cite{CPPACS01}.}
\label{fig:EOS}
\end{minipage}
\hfill
\begin{minipage}[t]{78mm}
\centerline{\epsfxsize=7.6cm \epsfbox{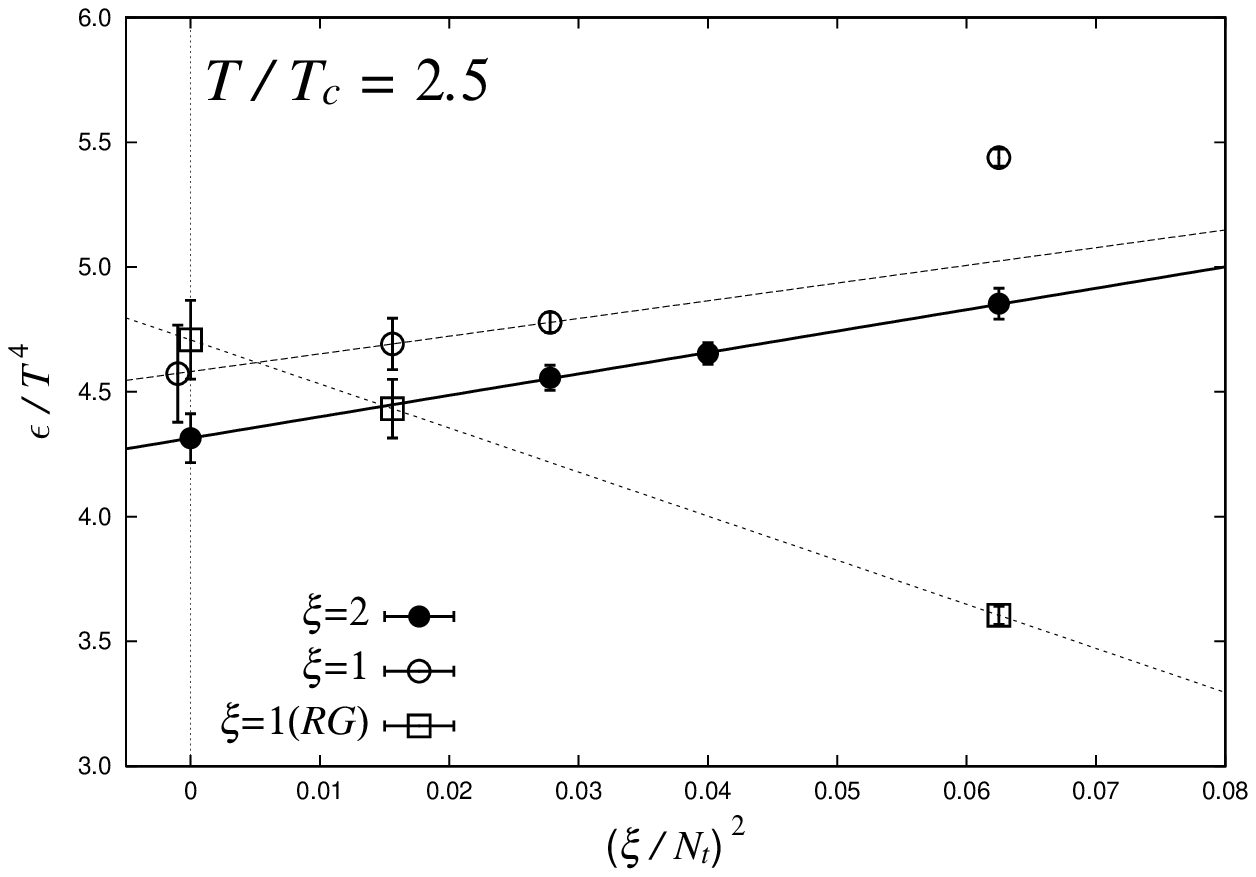}}
\vspace{-0.9cm}
\caption{Continuum extrapolation of the energy density at $T/T_c=2.5$ in SU(3) gauge theory on isotropic ($\xi=1$) and anisotropic ($\xi=2$) lattices \protect\cite{Namekawa01}.
}
\label{fig:aniso}
\end{minipage}
\end{figure}

As mentioned in the previous section, improvements of lattice actions are effective for reducing lattice artifacts at small $N_t$. 
For the EOS, however, improvement seems to be insufficient at $N_t=4$.
Fig.~\ref{fig:EOS} shows the energy density in QCD with Wilson-type quarks using improved actions (a renormalization-group improved gauge action combined with a clover-improved Wilson quark action) \cite{CPPACS01}. 
The discrepancies between the results at $N_t=4$ and 6 are as large as 40--50\%, which implies that the $N_t=4$ values suffer from large lattice errors.
The discrepancies are too large to attempt a continuum extrapolation $N_t \rightarrow \infty$ using data at $N_t=4$ and 6. 

On the other hand, the results obtained on $N_t=6$ lattices were suggested to be close to the continuum limit \cite{CPPACS01}, because the EOS from the staggered quark action \cite{Ber97b} are quite similar on $N_t=6$ lattices.
Thus, a precise continuum extrapolation may be possible if additional data at $N_t=8$ are generated.%
\footnote{
Here, it is worth to note that a ratio of observables, such as the Wroblewski parameter $\lambda_s \approx \chi_s/\chi_u$ \cite{Gavai02}, can be less sensitive to lattice errors.
For these quantities, accurate estimations may be possible even at small $N_t$. 
}

The recent study by the MILC Collaboration \cite{MILCLat02} using improved staggered-type quarks seems to be supporting this expectation.
Although the EOS has not been calculated, their results for triplet quark number susceptibility for $N_F=3$ clearly show that the $N_t=6$ and 8 data are roughly consistent with each other while $N_t=4$ data clearly deviates from them (Fig.~\ref{fig:2+1}(b)).

Calculation of the EOS at $N_t\geq8$ is, however, quite expensive.
A possible solution of the problem was proposed in Ref.~\cite{Namekawa01}.
The idea is as follows: a study of thermodynamic quantities in the high temperature limit shows that the leading lattice errors from the temporal lattice cut-off are larger than those from the spatial cut-off; therefore, the lattice artifacts in these quantities may be efficiently reduced on anisotropic lattices with $a_t < a_s$.
From this study, $\xi\equiv a_s/a_t = 2$ is found to be optimum.

The idea was tested in quenched QCD \cite{Namekawa01}.
In Fig.~\ref{fig:aniso}, the energy densities obtained on isotropic (open circles) and anisotropic (filled circles) lattices using the standard plaquette action are shown together with that on isotropic lattices using a renormalization-group improved action (open squares).
Results of continuum extrapolations, assuming the leading $1/N_t^2$ scaling law, are also given by corresponding symbols at $\xi/N_t=0$.
The right-most open circle, obtained on an $N_t=4$ isotropic lattice, shows a large deviation from the continuum limit, and, simultaneously, deviates from the leading scaling line. 
Therefore, the continuum extrapolation had to be made using only two data points at $N_t=6$ and 8 excluding $N_t=4$.
The filled circles are the results of anisotropic lattices.
The advantage of using anisotropic lattices is apparent.
On the coarsest lattice $N_t/\xi=4$, finite lattice spacing error at $\xi=2$ is much smaller than that at $\xi=1$ with the same plaquette action. 
Furthermore, even at $N_t/\xi=4$, the data is well described by the leading scaling. 
Therefore, we can confidently perform a continuum extrapolation using three data points, resulting in smaller and more reliable errors in the continuum limit.
Actually, this is the first well-controlled continuum extrapolation of the EOS in QCD even in the quenched approximation.
This anisotropic calculation of the EOS was about factor 5 more efficient in computational cost than isotropic calculations. 

Anisotropic lattices may help also in QCD with dynamical quarks.
Here, improvement of lattice actions is also important.
A status report toward this direction was presented by Umeda at Lattice 2002 \cite{Umeda01}.

%%%%%%%%%%%%%%%%%%%%%%%%%%%%%%%%%%%%%%%%%%%%%%%%%%%%%%%
\subsection{Spectral functions by MEM}
\label{sec:MEM}

Hadronic spectral functions are important in various phenomenological analyses.
It is desirable to determine them on the lattice directly from the first principles of QCD.
In Euclidian space-time, the spectral function $f$ is related to 
a two-point function $D$ by a kind of Laplace transformation:
\be
D(\tau) = \langle 0 | O(\tau) O^\dagger(0) | 0 \rangle 
 = \int_0^\infty d\omega K(\omega,\tau) f(\omega), 
\;\;\;\; K(\omega,\tau) = e^{-\omega\tau} + e^{-\omega(T^{-1}-\tau)},
\label{eq:D}
\ee
where $O$ is a hadronic operator and $T^{-1} = N_t a_t$.
However, because only a finite number of discrete data points are available for $D$, determination of the continuous $f$ from $D$ is an ill-posed problem.

Recently, a new method, the maximum entropy method (MEM), has been started to be applied to calculate $f$ from the lattice data.
MEM has been successfully applied to solve similar ill-posed problems in other fields, such as the image reconstruction problem in astrophysics or detective works.

In MEM, we calculate the ``most probable'' $f$ by maximizing the probability $P[f|DH]$ of $f$ when the data $D$ and prior knowledge $H$ on $f$ are given. 
By Bayes' theorem, we can decompose $P[f|DH]$ as $P[f|DH] \propto P[D|fH] \times P[f|H]$. 
The first part is the probability of $D$ given $f$ and $H$, and can be evaluated by the conventional least $\chi^2$ likelihood function $L$ as $P[D|fH] \propto e^{-L[D,f]}$. 
The second part $P[f|H]$, the probability of $f$ when $H$ is given, is something only God knows and concentrates the information what we think ``probable''.
In MEM, the Shannon-Jaynes entropy $S$ is adopted to evaluate 
\be
P[f|H] \propto e^{\alpha \, S[f,m]}, \;\;\;\;
S = \int_0^\infty d\omega \left\{ f(\omega) - m(\omega) - f(\omega) \ln \left[f(\omega)/m(\omega)\right] \right\},
\label{eq:SJ}
\ee
with $m(\omega)>0$ a real function called the default model and $\alpha$ a positive parameter to specify the relative weight between $L$ and $S$.
(Note that $L$ attracts $f$ to $D$ while $S$ attracts $f$ to $m$.)
Eq.(\ref{eq:SJ}) means that we use the term ``probable'' in the sense of the information theory.
Here, two additional inputs, $\alpha$ and $m(\omega)$, are required.
The $\alpha$-dependence turns out to be weak in most cases and can be eliminated by an additional statistical argument. 
The $m$-dependence is also (usually) weak. 

From various tests on the lattice, the method has been shown to work well provided that the number of data is large and the statistical quality is high. 
MEM can correctly reproduce the ground state and the first exited state energies from the location of peaks in $f$, as well as the decay constants for these states from the peak area \cite{Asakawa99,Yamazaki01}.
When $D$ at small distances are included, MEM is accurate enough to reproduce lattice artifacts (doublers etc.) too \cite{Yamazaki01}.
It was also demonstrated in a sigma model that MEM is capable of describing decay modes \cite{YamazakiLat02}.
A benefit of MEM is that the tasks required to extract the information are much lighter than those with the conventional methods.

It should be nevertheless stressed that, no matter how plausibility arguments are accumulated, answers to an ill-posed problem are always heuristic.
Therefore, it is important to understand which properties of $f$ are reliably calculated. 
For that, let us reexamine the role of the input data.
From Eq.(\ref{eq:D}), we see that $D$ is essentially the area of $f$ with an exponential cut-off at $\omega \sim 1/\tau$.
Therefore, we can draw following conclusions: 
(1) When we remove $D$ at $\tau<\tau_{min}$ to avoid lattice artifacts, $f$ at $\omega$ \simgt $c/\tau_{min}$ cannot be determined. 
In accord, the choice of $m$ was found to affect $f$ at large $\omega$  \cite{Asakawa99}.
Here, $c = O(1)$ is a constant depending on the precision of $D$ and the required confidence level on $f$.
(A crude estimation leads $c \sim 2$ (8) for the cases of 10\% (2\%) errors in $D$ at $\tau \sim 1/2T$, assuming the Born form of $f$ \cite{KarschDilepton}.)
(2) $f$ at $\omega << 1/\tau_{max}$ is also insensitive to the data. 
Actually, when $\tau_{max}$ is not sufficiently large, we sometimes encounter false peaks at small $\omega$ \cite{Yamazaki01}. 
See also arguments in \cite{Aarts02}.
(3) The location and the area of a peak will be reliable when we input precise data at around $\tau \sim 1/\omega_{\rm peak}$.
(4) To extract more detailed features of the peak, such as the width, we need enough number of precise data points within the corresponding range of $\tau$.

As stressed in \cite{Asakawa99}, a rather large number of data points with high statistics is required to calculate $f$ precisely. 
For example, in a study of finite-temperature QCD, $N_t>30$ was argued to be necessary. \cite{AsakawaQM02}. 
Because a simulation at large $N_t$ is demanding, it is natural to adopt anisotropic lattices. 
In \cite{AsakawaQM02}, a lattice with $\xi=4$ was studied.

\begin{figure}[t]
\begin{minipage}[t]{76mm}
\centerline{\epsfxsize=7.4cm \epsfbox{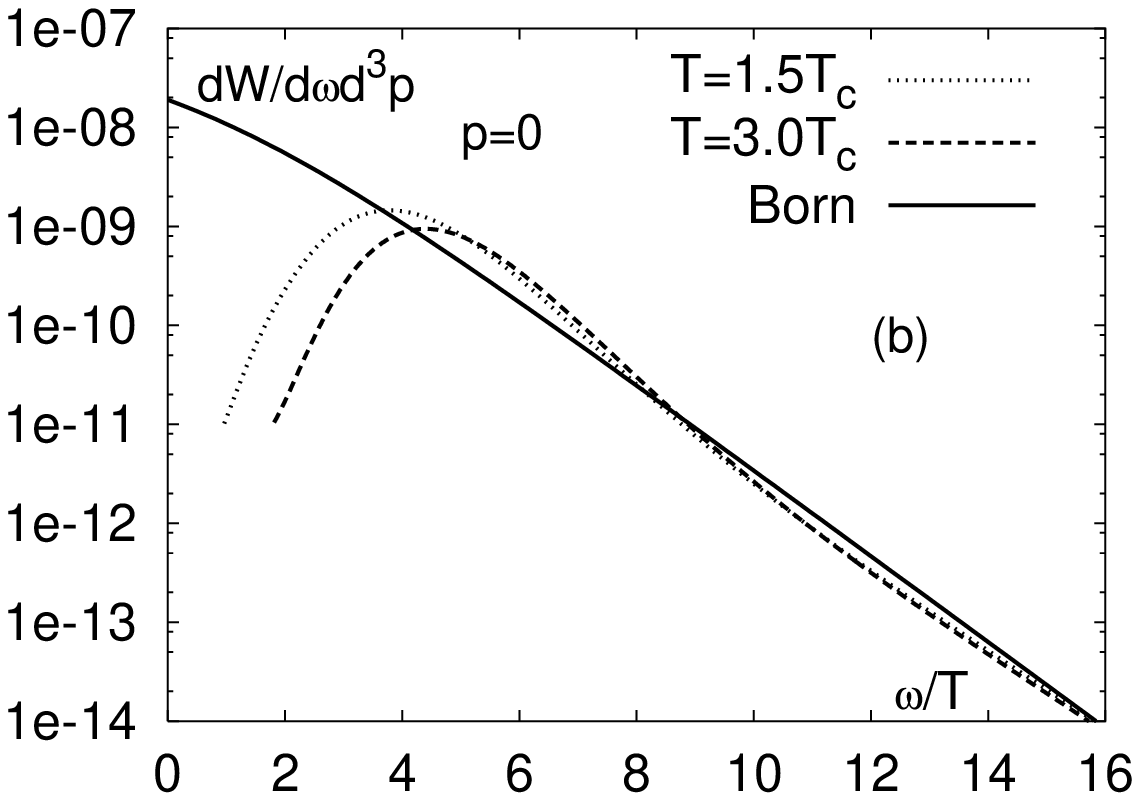}}
\vspace{-0.8cm}
\caption{Dilepton rate from vector spectral function in quenched QCD \protect\cite{KarschDilepton}.}
\label{fig:dilepton}
\end{minipage}
\hfill
\begin{minipage}[t]{76mm}
\centerline{\epsfxsize=7.4cm \epsfbox{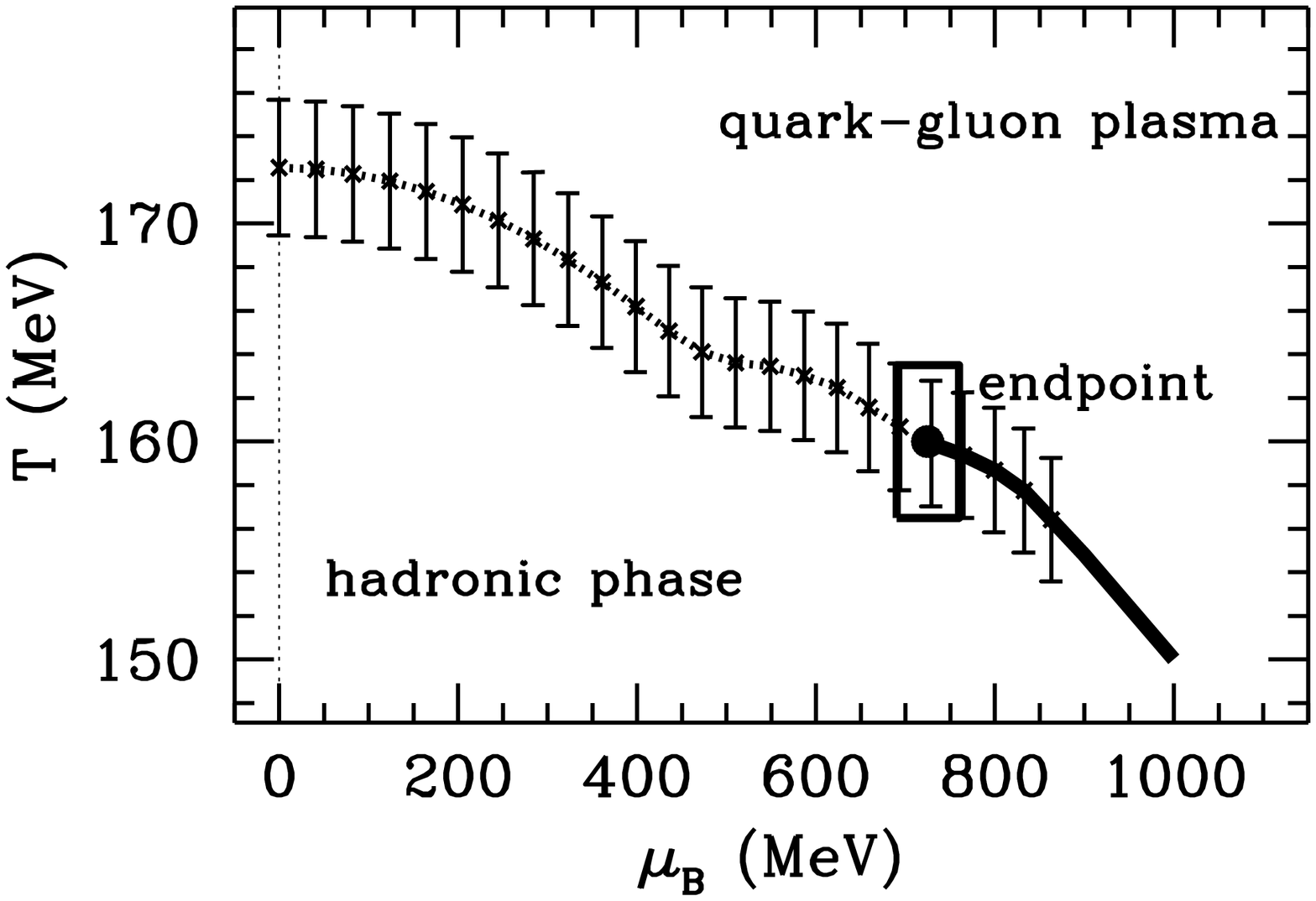}}
\vspace{-0.8cm}
\caption{Phase diagram obtained by reweighting from $T_c$ at $\mu=0$ \protect\cite{FK}.}
\label{fig:FK}
\end{minipage}
\end{figure}

Applications of MEM to phenomenological quantities have already been started in the quenched approximation of QCD. 
In \cite{KarschDilepton}, thermal dilepton rate was calculated from a vector spectral function on isotropic $N_t=12$ and 16 lattices.
From the observed suppression of $f$ at small $\omega$, they suggested decrease of the dilepton rate relative to the Born value at small $\omega$ (see Fig.~\ref{fig:dilepton}).
A high number of data points and good statistics seem to be critical if a decisive conclusion is to be drawn; however, because a recent higher quality data \cite{AsakawaQM02} also supports the behavior of $f$ at small $\omega$, the decrease of the dilepton rate at small $\omega$ seems to be a real effect.

%%%%%%%%%%%%%%%%%%%%%%%%%%%%%%%%%%%%%%%%%%%%%%%%%%%%%%%%%%%%%%%%%%%%%
\section{FINITE DENSITY QCD}
\label{sec:mu}

Rich physical content is expected in finite density QCD both theoretically and experimentally.
However, a numerical simulation of QCD at finite chemical potential $\mu$\ has remained to be one of the most challenging issues since the first foundation of lattice QCD, because, in the path integral of QCD
\be
Z = \int dU \, e^{-S_{gauge}}\, \det M (\mu),
\label{eq:QCDZ}
\ee
the quark determinant $\det M$ becomes complex at $\mu\neq0$ even in Euclidian space-time: $\det M = \left| \det M \right| \, e^{i\theta}$.

A trick to validate the statistical treatment of Eq.(\ref{eq:QCDZ}) in Monte Carlo simulations is to regard the phase factor $e^{i\theta}$ as a part of observables.
However, the expectation value of the phase factor is found to be exponentially suppressed by the lattice volume, $\langle e^{i\theta} \rangle \propto e^{-V}$.
Therefore, exponentially large statistics is required as we increase $V$ to keep the accuracy of the results. 
This difficulty is called the phase (sign) problem.

%%%%%%%%%%%%%%%%%%%%%%%%%%%%%%%%%%%%%%%%%%%%%%%%%%%%%%%
\subsection{QCD at small chemical potentials}
\label{sec:smallmu}

Recently, a (partial) breakthrough of the problem was introduced by Fodor and Katz \cite{FK} applying the reweighting method (spectral density method) to study QCD at small densities. 
Basic idea of the reweighting is simple \cite{reweighting}.
Because the functional form of the measure is known in Eq.(\ref{eq:QCDZ}), we can correct the distribution of operators measured at a point $A$ in the coupling parameter space to that at any other point $B$ as
\be
\underbrace{d U e^{-S_{gauge}(B)} \det M(B)}_{\rm target \; point} 
=
\underbrace{d U e^{-S_{gauge}(A)} \det M(A)}_{\rm simulation \; point} 
\times
\underbrace{e^{S_{gauge}(A)-S_{gauge}(B)} \frac{\det M(B)}{\det M(A)}}_{\rm correction \; factor} 
\label{eq:reweighting}
\ee
In practice, however, to obtain a reliable distribution at $B$, the distributions at $A$ and $B$ should have a reasonable overlap.

Reweighting from $\mu=0$ (the Glasgow method) is a natural strategy to avoid simulations at $\mu\ne0$.
Since early 90's, however, no attempts made at $T=0$ have been able to produce a sensible result even at very small $\mu$. 
We understand that the failure is caused by exponentially small width of the distribution in the $\mu$-direction due to the phase problem. 

Here, we can imagine several special cases where we may expect wider distributions:
(i) Small system volume $V$. 
(ii) High temperature. 
(iii) Phase transition/crossover point. 
These factors will have a cooperative effect. 

Based on a deep insight of the problem, Fodor and Katz combined all the three and, for the first time, succeeded to obtain a reasonable result for QCD at $\mu\ne0$ \cite{FK}.
On $N_s^3\times4$ lattices with $N_s=4$, 6 and 8, they performed a simulation at the finite-temperature transition point of $\mu=0$ QCD, and looked for the points with finite overlap in the $(\mu,T)$ plane as shown in Fig.~\ref{fig:FK}.
Among these points, they found the endpoint of the first order transition line between hadronic and QGP phases at $T_E=160\pm3.5$ MeV and $\mu_E=725\pm35$ MeV.

Because the simulation is done with heavy staggered quarks ($m_{ud} a = 0.025$ and $m_s a=0.2$) using the standard action, the finite temperature transition is crossover at $\mu=0$. 
%The points with finite overlap with the simulation point should be also transition or crossover between confining and deconfining phases.
When we decrease the quark masses toward the physical point, the end point of first order transition line will move toward smaller $\mu$.
As discussed in Sec.~\ref{sec:3flavor}, unimproved lattice actions suffer from sizable lattice artifacts at $N_t=4$.
Whether the first order line reaches the $T$-axis for the physical quark masses is currently unclear.

In QCD with dynamical quarks, evaluation of the determinants in Eq.(\ref{eq:reweighting}) is quite time-consuming on large lattices. 
Recently, Allton et al.\ applied a Taylor expansion method to calculate $\det M$ at small $\mu$ \cite{Allton02}, 
\be
\ln \det M(\mu) = \ln \det M(0) + {\rm Tr} \left( M^{-1} \frac{\partial M}{\partial \mu}\right)_{\mu=0} \, \mu + \, \cdots.
\ee
When we adopt the noise method to calculate the quark traces, large lattice volume is not a serious problem at least in this part of the calculation.
Accordingly, they performed a simulation on a $16^3\times 4$ lattice in 2 and 3-flavor QCD with improved staggered quarks.
They confirmed that the method applies well to the RHIC region ($\mu/T \approx 0.1$), and calculated several thermodynamic quantities.
In particular, the effect of $\mu$ in the EOS is estimated to be O(1\%) at the RHIC point.

Another approach is to try an analytic continuation from pure imaginary $\mu$ where $\det M$ is real \cite{Forcrand02} (see also \cite{DElia02,Takaishi02}).
This method is also limited to small real $\mu$ because (i) $|{\rm Im}\mu| < \pi T/3$ due to a Z(3) periodicity in the imaginary $\mu$ direction and (ii) the number and statistical accuracy of data are limited.
Concretely, we first fit the data at several values of imaginary $\mu$ by a low-order polynomial in $\mu$, and replace $\mu$ by $-i\mu$ in the final fit-function. 
Results on $6^3\times4$ and $8^3\times4$ for $N_F=2$ QCD are reported to be consistent with those from other groups.

%%%%%%%%%%%%%%%%%%%%%%%%%%%%%%%%%%%%%%%%%%%%%%%%%%%%%%%
\subsection{Large chemical potentials}
\label{sec:largemu}

QCD was suggested to have rich phase structures at high densities \cite{ColorSuper}.
The region in the phase diagram may be studied by GSI and JHF. 
The methods described in Sec.~\ref{sec:smallmu}, however, cannot be applied to study the high-density regions.
To obtain insights for QCD at large $\mu$, a special case of QCD and several QCD-like theories are studied in which $\det M$ is real even at $\mu\ne0$. 

When we consider QCD with isovector chemical potential, $\mu_u=-\mu_d$, the functional measure $[\det M(\mu) \det M(-\mu)]$ is real because $\det M(-\mu) = (\det M(\mu))^*$.
Simulations indicate pieces of evidence for the expected phase transition to a phase with $\pi^+$-condensation.
Results of observables are compatible with predictions from an effective chiral theory \cite{isovector}. 

The quark determinant is also real at $\mu\ne0$ in two-color QCD \cite{KogutSU2,Hands01,GavaiSU2,Muroya02}.
From simulations, existence of di-quark condensation has been demonstrated and consistency with effective chiral theory has been reported.
Muroya et al.\ studied meson masses at $\mu\ne0$ and showed that $m_\rho$ decreases with increasing $\mu$ \cite{Muroya02}, which may explain the dilepton enhancement observed by CERES \cite{CERES} if the behavior remains in real QCD.

Finally, a simulation of the NJL model showed an evidence of a BCS transition in accord with the color-superconducting scenario \cite{HandsNJL}.
See also \cite{Miyamura02} for other study of NJL.

%%%%%%%%%%%%%%%%%%%%%%%%%%%%%%%%%%%%%%%%%%%%%%%%%%%%%%%%%%%%%%%%%%%%%
\section{CONCLUSIONS}
\label{sec:summary}

Since the QM 2001 conference, a couple of steps have been made in lattice QCD to reduce the distance between theory and experiment.
For several quantities, lattice studies of QCD at finite temperatures are entering the stage of precise quantitative predictions. 

New 2+1 flavor simulations have further constrained the finite-temperature QCD transition, although the clarification of the nature of the transition at the physical point has remained an open problem. 
Simulations at large $N_t$ suggest that a precise continuum extrapolation of thermodynamic quantities may be possible with data at $N_t=6$, 8, $\cdots$.
Calculation of EOS at $N_t \ge 8$ is, however, demanding. 
A way out will be given by anisotropic lattices, which have enabled us to carry out the first well-controlled continuum extrapolation of the EOS in quenched QCD. 
An extension to QCD with dynamical quarks is currently being attempted.
The MEM will enlarge the predictive power of lattice when used carefully, and the first applications to the quark matter phenomenology have already begun.
The hard numbers, such as $T_c^{(N_f=2)} \approx 170$--175 MeV, 
$\epsilon^{(N_F=2)} / T^4 |_{T/T_c=1.5} \approx 14$,
$\epsilon^{(N_F=2)} / T^4 |_{T/T_c=1.5} \approx 3$, are unchanged through these developments, but the confidence in these values has become much firmer.

In the last decades, the finite density QCD has been a barren area in lattice QCD in spite of its great importance and big efforts, due to a technical difficulty of numerical simulations at finite $\mu$.
Last two years, however, a (partial) breakthrough has been made for the case of small chemical potentials around the finite temperature transition point, and the first sensible results for QCD at $\mu\ne0$ are obtained.
Different groups, adopting different methods, have reached a general agreement of results.
Further studies are underway to reduce the lattice artifacts.
These methods cannot be simply applied for large $\mu$. 
To obtain insights for QCD at large $\mu$, several QCD-like theories, for which the conventional technique is applicable also at large $\mu$, are being studied.
Results obtained so far seem to be consistent with the color-superconducting scenario.

%%Acknowledgements%%

I thank M.\ Asakawa, S.\ Ejiri, Z.\ Fodor, R.V.\ Gavai, F.\ Karsch, V.I.\ Lesk, Y.\ Miake, A.\ Nakamura, Ch.\ Schmidt, R.\ Sugar and the members of the CP-PACS Collaboration for valuable discussions, comments and data.
This work is supported in part by Grants-in-Aid of the Ministry of Education (Nos.\ 12304011 and 13640260).

%%%%%%%%%%%%%%%%%%%%%%%%%%%%%%%%%%%%%%%%%%%%%%%%%%%%%%%%%%%%%%%%%%%%%

\end{document}